\documentclass{aa}
\usepackage{graphicx}
\usepackage{txfonts}
\usepackage{siunitx}
\usepackage{hyperref}
\usepackage{lipsum}
\usepackage{bm}
\usepackage{xcolor}
\usepackage{lipsum}
\usepackage{placeins}

\usepackage{verbatim}

\newcommand\iona[2]{#1\,{\sc #2}}

\usepackage{xpatch}
\usepackage{hyperref}
\hypersetup{
colorlinks=true,
breaklinks=true,
citecolor=blue,
allcolors=blue,
frenchlinks=true
}

\makeatletter
\xpatchcmd\NAT@citex
{%
\@citea\NAT@hyper@{%
\NAT@nmfmt{\NAT@nm}%
\hyper@natlinkbreak{\NAT@aysep\NAT@spacechar}{\@citeb\@extra@b@citeb}%
\NAT@date
}%
}
{%
\@citea
\NAT@nmfmt{\NAT@nm}%
\NAT@aysep\NAT@spacechar
\NAT@hyper@{\NAT@date}%
}
{}{}
\xpatchcmd\NAT@citex
{%
\@citea\NAT@hyper@{%
\NAT@nmfmt{\NAT@nm}%
\hyper@natlinkbreak{\NAT@spacechar\NAT@@open\if*#1*\else#1\NAT@spacechar\fi}%
{\@citeb\@extra@b@citeb}%
\NAT@date
}%
}
{
\@citea
\NAT@nmfmt{\NAT@nm}%
\NAT@spacechar\NAT@@open\if*#1*\else#1\NAT@spacechar\fi
\NAT@hyper@{\NAT@date}%
}
{}{}
\makeatother

\makeatletter
\renewcommand*\aa@pageof{, page \thepage{} of \pageref*{LastPage}}
\makeatother

\begin{document}

\title{Comparing the interstellar and circumgalactic origin of gas in the tails of jellyfish galaxies}
\titlerunning{On the tails of jellyfish galaxies}

\author{Martin Sparre\inst{1,2}
\and
Christoph Pfrommer\inst{2}
\and
Ewald Puchwein\inst{2}
}

\institute{Institut f\"ur Physik und Astronomie, Universit\"at Potsdam, Karl-Liebknecht-Str.\,24/25, 14476 Golm, Germany
\and
Leibniz-Institut f\"ur Astrophysik Potsdam (AIP), An der Sternwarte 16, 14482 Potsdam, Germany
}

\date{\today}


\abstract{
  Simulations and observations find long tails in `jellyfish galaxies', which are commonly thought to originate from ram-pressure stripped gas of the interstellar medium (ISM) in the immediate galactic wake. While at larger distances from the galaxy, they have been claimed to form in situ owing to thermal instability and fast radiative cooling of mixed ISM and intracluster medium (ICM). 
  In this paper, we use magnetohydrodynamical windtunnel simulations of a galaxy with the {\sc arepo} code to study the origin of gas in the tails of jellyfish galaxies. To this end, we model the galaxy orbit in a cluster by accounting for a time-varying galaxy velocity, ICM density and turbulent magnetic field. Tracking gas flows between the ISM, the circumgalactic medium (CGM) and the ICM, we find contrary to popular opinion that the majority of gas in the tail originated in the CGM. 
  Prior to the central passage of the jellyfish galaxy in the cluster, the CGM is directly transported to the clumpy jellyfish tail that has been shattered into small cloudlets. After the central cluster passage, gas in the tail originates both from the initial ISM and the CGM, but that from the latter was accreted to the galactic ISM before being ram-pressure stripped to form filamentary tentacles in the tail. Our simulation shows a declining gas metallicity in the tail as a function of downstream distance from the galaxy. 
We conclude that the CGM plays an important role in shaping the tails of jellyfish galaxies.
}


\maketitle
%

\section{Introduction}

The high density of galaxies in clusters and the existence of a hot X-ray emitting intracluster medium (ICM) play an important role in transforming the morphology and star formation properties of cluster galaxies, which is manifested in the morphology-density relation \citep{1980ApJ...236..351D}: at the present day, galaxy clusters host larger fractions of galaxies that are morphologically classified as ellipticals and lenticular galaxies of type S0. In addition, cluster galaxies are on average redder, are more massive, more concentrated, less gas rich and have lower specific star formation rates in comparison to low-density environments in the `field' where spiral galaxies dominate the numbers. When a spiral galaxy falls into a cluster, it is exposed to the ram pressure provided by the hot ICM that sweeps clean the  interstellar medium (ISM) as pointed out by \citet{1972ApJ...176....1G}. In dense environments, this effect may be responsible for transforming actively star forming disc galaxies into red S0 galaxies \citep{2000Sci...288.1617Q} and dominates over tidal effects which affect both gas and stars and often lead to galaxy harassment \citep{1996Natur.379..613M}.

Three-dimensional hydrodynamical simulations demonstrate that the properties of galaxies are significantly impacted if they are exposed to a ram-pressure wind \citep[e.g.][]{2001MNRAS.328..185S, 2005A&A...433..875R,2012MNRAS.422.1609T}. As the ISM is stripped from the galactic disc, the outer disc is truncated while the inner disc is compressed, which results in the emergence of numerous flocculent spiral arms. This compression of the central ISM can instigate intense star formation within the first $10^8$ years of interaction \citep{2001MNRAS.328..185S, 2003ApJ...596L..13B}, even though this star formation activity is expected to diminish over time as the galactic ISM is exhausted. This initial burst of star formation may elucidate the presence of blue `Butcher-Oemler galaxies' observed in $z \gtrsim 0.3$ clusters \citep{1984ApJ...285..426B}. Depending on the properties of magnetic fields, thermal conduction and viscosity, it is possible that some of the stripped ISM could cool sufficiently to form stars, particularly since the primary heating source of the ISM, stellar UV radiation, is significantly weaker in intracluster space. Indeed, magnetic fields are draped over these galaxies as they orbit the magnetised ICM to form a protective layer that increases the stripping time-scale and shields the stripped tails from the hot ICM wind \citep{2008ApJ...677..993D, 2010NatPh...6..520P, 2014ApJ...784...75R,2020MNRAS.499.4261S}. This naturally explains the large-scale magnetic field ($\gtrsim4~\mu$G) and extremely high fractional polarisation ($>50 \%$) in the long H$\alpha$-emitting tail of the jellyfish galaxy JO206 \citep{2021NatAs...5..159M}.

Simulating ram-pressure stripping of disc galaxies that are embedded in their circumgalactic medium (CGM), enabled \citet{2009MNRAS.399.2221B} to coin the term `jellyfish galaxy' because of the morphological similarity of the stripped CGM of a galaxy and a jellyfish. High-resolution hydrodynamical simulations of ram pressure-stripped galaxies show little difference in the amount of stripped gas between the case that self-consistently produces a clumpy, multi-phase ISM and a comparison simulation without radiative cooling. However, in the multi-phase galaxy, the gas is stripped more rapidly and to a smaller radius \citep{2009ApJ...694..789T}. Radiation-hydrodynamic simulations of gas-rich dwarf galaxies with a multi-phase ISM show that the ram-pressure stripped ISM is the primary origin of molecular clumps in the immediate wake within 10 kpc from the galactic plane, whereas in situ formation owing to ISM-ICM mixing and cooling predominates as the primary mechanism for dense gas in the distant tail of the galaxy \citep{2020ApJ...905...31L,2022ApJ...928..144L}. Indeed, the mixing scenario is facilitated for Milky Way–like galaxies exposed to a high-density and low-velocity ICM wind as shown in hydrodynamical simulations \citep{2021ApJ...911...68T}, which implies a decreasing metallicity profile with distance along the tail as found in Multi Unit Spectroscopic Explorer (MUSE) observation \citep{2021ApJ...922L...6F}.

Observations of individual jellyfish galaxies show `tentacle-like' tails containing cold gas detectable in \iona{H}{I} \citep{2019MNRAS.487.4580R} and molecular hydrogen that are often coincident with H$\alpha$ \citep{2008ApJ...688..918Y, 2017ApJ...844...48P} and X-ray emission \citep{2006AJ....131.1974O, 2007ApJ...671..190S}. This demonstrates the true multi-phase nature of these tails with enshrouded sites of star formation \citep[for a review, see][]{2022A&ARv..30....3B}. All these authors conclude that ram pressure stripping of the galactic ISM is responsible for the tails, while heat conduction and advection from the hot ICM may also contribute to powering the optical lines \citep{2007ApJ...671..190S} and X-ray tail \citep{2019ApJ...887..155P}.

To answer the question about the ubiquity of this jellyfish phenomenon and its role in galaxy evolution in dense environments, progressively larger sample sizes have been constructed. More than 10 asymmetric star forming galaxies have been found in the outskirts of Coma, showing gas stripping from their discs and sometimes long tails reaching 100 kpc \citep{2010MNRAS.408.1417S, 2010AJ....140.1814Y}. An analysis of several hundred  low-redshift jellyfish galaxies with various asymmetric and disturbed morphologies show signs of triggered star formation \citep{2016AJ....151...78P}. This was the basis for a new integral-field spectroscopic survey with MUSE at the Very Large Telescope called GAs Stripping Phenomena in galaxies with MUSE (GASP), which studies gas removal processes in galaxies and compares jellyfish galaxies to a control sample of disc galaxies with no morphological anomalies \citep{2017ApJ...844...48P}. Interestingly, the jellyfish phenomenon does not only affect a few cluster galaxies but instead several tens of galaxies, provided the data is sufficiently deep \citep{2019MNRAS.484..892R, 2021A&A...648A..63D}.

This progress in observations calls for a significantly improved simulation modelling that accounts for (i) a time varying density and velocity along the galaxy orbit in a cluster \citep[similar to][]{2016A&A...591A..51S}, (ii) a time varying turbulent magnetic field to enable the process of magnetic draping, and (iii) it requires the inclusion of the CGM of the galaxies in addition to the ISM in the disc. The CGM represents a reservoir of accreted gas from the cosmological surroundings and out of which gas condenses onto the disc providing fuel for ongoing star formation, and into which galactic winds driven by stellar feedback and active galactic nuclei deposit baryons from the disc \citep{2017MNRAS.470.4698A,2019MNRAS.483.4040S, 2020ARA&A..58..363P}. It contains a significant fraction of the halo's baryon budget \citep{2017ARA&A..55..389T}. It is multi-phase, which is observationally established by detection of various ions probing different temperature--density regimes in absorption sight-lines \citep{2013ApJS..204...17W,2016ApJ...833...54W,2017ApJ...837..169P,2017A&A...607A..48R}. Idealised hydrodynamical simulations \citep{2017ApJ...834..144S,2018ApJ...862...56S,2021MNRAS.505.1083G} and cosmological simulations \citep[e.g.][]{2019MNRAS.482L..85V} also favour the CGM to be multi-phase since clouds larger than the cooling length are thermally unstable \citep{2018MNRAS.473.5407M,2019MNRAS.482.5401S}. Furthermore, cold gas is expected to be able to survive and grow in certain halo environments \citep{2017MNRAS.470..114A,2018MNRAS.480L.111G,2020MNRAS.492.1841L,2020MNRAS.499.4261S,2021MNRAS.501.1143K,2022ApJ...925..199A,2023arXiv230703228A}.

In this paper, we critically scrutinise the common wisdom, that jellyfish tails are primarily produced by ram-pressure stripping of disc ISM which is complemented by mixing of ISM and ICM gas at larger distances from the galaxy, and we specifically determine the role of CGM gas in the tail. To this end, we follow the galaxy orbit in a cluster by accounting for a time-varying galaxy velocity, an ICM density, and a turbulent magnetic field in a versatile and flexible windtunnel setup. This allows us to vary important parameters in a controlled fashion and to reach very high resolution. Our approach is complementary to cosmological simulations, which are ideally suited for studying the statistics of orbits of jellyfish galaxies and include a cosmologically self-consistent CGM, while these simulations have to make compromises with numerical resolution. The Illustris TNG simulation, for example, has many jellyfish galaxies, where stripping is ongoing \citep{2019MNRAS.483.1042Y}. In a more recent simulation, Illustris TNG50, ram-pressure stripping is also found to be the main mechanism causing gas removal from jellyfish galaxies in halos with virial masses of $10^{12-14.3}$ M$_\odot$ \citep{2023MNRAS.524.3502R}. In a Milky-Way-mass setting, satellites orbiting the halo CGM may also experience ram-pressure stripping \citep{2024arXiv240400129Z,2024MNRAS.527..265R}, and these are hence analogous to jellyfish galaxies in more massive galaxy clusters.

The outline of the paper is as follows. After laying down our method in Section~\ref{sec:methods}, we present our results on the origin of gas and the metallicity in the gaseous tails in Section~\ref{Sec:Results}. We discuss our results in Section~\ref{Sec:Discussion} and conclude in Section~\ref{Sec:Conclusion}. In Appendix~\ref{appendix1}, we present our definitions of the ISM, CGM, and ICM in phase space while we show our procedure of separating HI disc and tail in Appendix~\ref{AppendixRHI}.


\begin{figure}
\centering
\includegraphics[width=\linewidth]{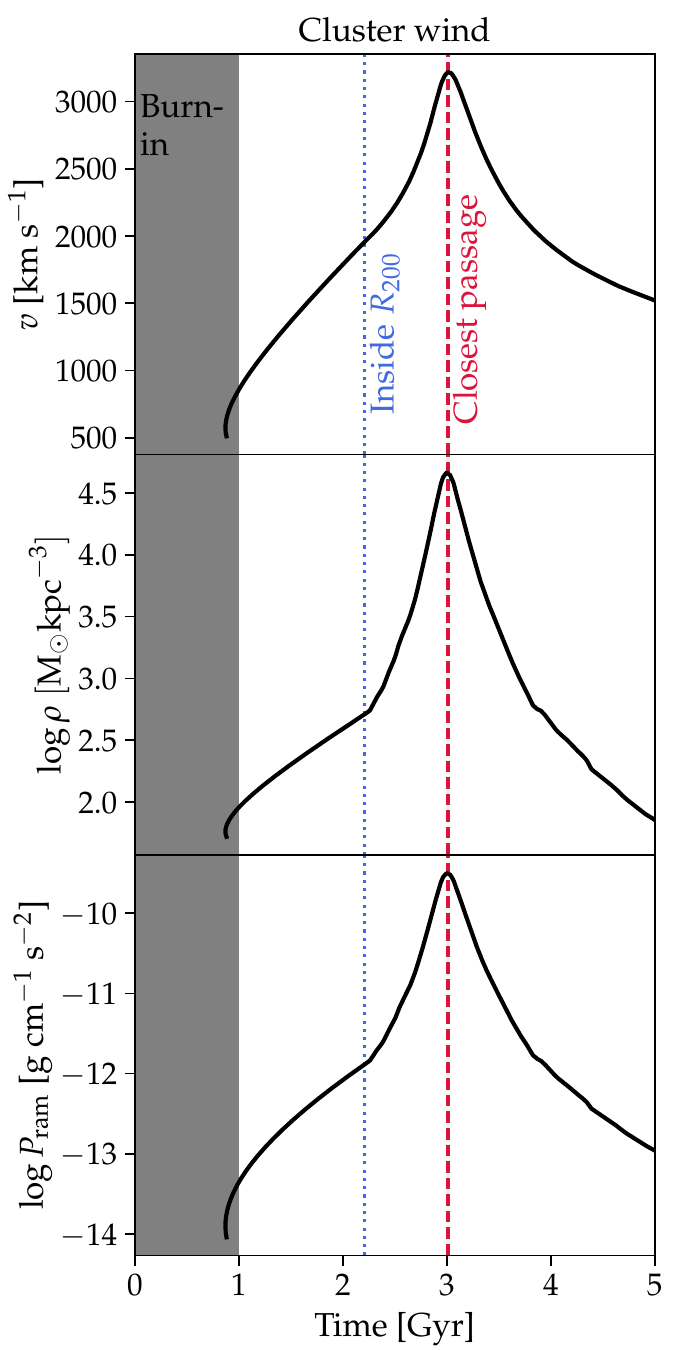}
\caption{Our magnetohydrodynamical windtunnel simulation uses a time-dependent wind mimicking what a jellyfish galaxy experiences while orbiting a $1.1\times 10^{15}$ M$_\odot$ cluster. This figure shows the resulting ICM wind speed, ICM wind density, and ram-pressure. In the first 1 Gyr we have a `burn-in phase', where the galaxy settles into an equilibrium. Shortly after 2.0 Gyr the galaxy crosses $R_{200}$ of the cluster, and 3.0 Gyr marks the closest passage to the centre, where the ram-pressure peaks.}
\label{Fig10_PlotClusterWindForPaper2}
\end{figure}

\section{Methods}
\label{sec:methods}

\subsection{Jellyfish galaxy simulation}

The initial conditions and simulation setup is summarised in \citet{2023arXiv231105679S}, where the jellyfish galaxy is initialised with a halo mass of $M_{200} = 2 \times 10^{12}$ M$_\odot$ and a disc with a stellar mass of $M_\star = 7.38 \times 10^{10}$ M$_\odot$. The galaxy has a hot CGM-like halo with a density profile following the same shape as the dark matter (a Hernquist profile, \citealt{1990ApJ...356..359H}) and a temperature that ensures pressure equilibrium of the gas. We use \textsc{makenewdisk} to create the initial galaxy \citep{2005MNRAS.361..776S}. In our initial conditions, the disc and CGM have a mass fraction ($m_\text{d}$ and $m_\text{HG}$, respectively, from Table~1 of \citealt{2023arXiv231105679S}) of 4.1 and 11.7 per cent of $M_{200}$. In this paper we analyse the high-resolution simulation, where the disc normal and the direction of the ICM wind form an angle of 60$^\circ$, which is an oblique configuration that is closer oriented to an edge-on inclination (simulation {\tt 60deg-Bturb-HR} from \citealt{2023arXiv231105679S}). This simulation has a baryonic mass resolution of $3.95 \times 10^4$~M$_\odot$.

Accounting for the time-dependence of the hydrodynamical properties of the ICM wind is important when simulating the evolution of jellyfish galaxies \citep{2019ApJ...874..161T}. Our time-dependent ram-pressure, $P_\text{ram}\equiv\rho \varv^2$, exerted by the ICM wind on the galaxy in our windtunnel setup is shown in Fig.~\ref{Fig10_PlotClusterWindForPaper2}. Our setup hence describes a jellyfish galaxy moving supersonically in the ICM (as also found in the cosmological simulation of \citealt{2019MNRAS.483.1042Y}). We adopt a 1 Gyr long `burn-in' phase, in which the galaxy in the initial conditions settles into an equilibrium and the CGM is enriched with outflows from stellar winds. After 1 Gyr, we ramp up the density, temperature, and velocity of the wind experienced by the galaxy, which is determined from a realistic orbit in a $M_{200}=1.1\times 10^{15}$ M$_\odot$ cluster (halo A from \citealt{2016A&A...591A..51S}). In practice, our windtunnel works by injecting density, temperature, magnetic field, velocity, and metallicity (see fig.~1 in \citealt{2023arXiv231105679S}) in a `wind injection region' at the lower part of the simulation box at $y<46$ kpc.

\begin{table*}
\centering
\caption{Origin and evolutionary paths of gas ending up in clumps in the galaxy tail traced using Monte Carlo tracers.}
\begin{tabular}{ll}
\hline\hline
{\bf Tracer path} &  Definition \\
\hline
$\text{ISM}\to\text{clumps}$ & Gas, categorised as ISM at $t=1$ Gyr, residing in tail clumps at the time of analysis. \\
$\text{CGM}\to\text{clumps}$ & Gas, categorised as CGM at $t=1$ Gyr, residing in tail clumps at the time of analysis. \\
$\text{ICM}\to\text{clumps}$ & Gas, injected as ICM at $t\geq 1$ Gyr, residing in tail clumps. \\
\hline
$\text{CGM}\to \text{ISM}\to \text{clumps}$ & A subset of the category `$\text{CGM}\to\text{clumps}$', which passes through the galaxy's ISM.\\
$\text{ICM}\to \text{ISM}\to \text{clumps}$ & A subset of the category `$\text{ICM}\to\text{clumps}$', which passes through the galaxy's ISM. \\
\hline\hline
\end{tabular}
\tablefoot{The first three rows of paths are based on the properties of a tracer at $t=1$ Gyr and at the time/snapshot of the analysis. The two lower rows show paths with an intermediate phase in the galaxy's ISM, which means gas characterised as ISM and residing no more than a distance $1.1 R_\text{HI}$ downstream from the galaxy centre, where $R_\text{HI}$ is a measure of the HI extend of the gas disc.}
\label{Table:gastracers}
\end{table*}

We use a turbulent time-dependent magnetic field in the wind with a thermal-to-magnetic pressure ratio of 50. The ICM wind and CGM is initialised with a third of solar metallicity, and the ISM itself is initialised with solar metallicity. We use solar abundances.

The ram-pressure stripping condition from \citet{1972ApJ...176....1G} and \citet{2010gfe..book.....M} reads
\begin{align}
P_\text{ram} > 2\pi \Sigma_{\star\text{,\,disc}}\,\Sigma_\text{gas,\,disc}.\label{GunnGott}
\end{align}
As a result, the ISM of jellyfish galaxies are `outside-in stripped', because the right-hand-side of this equation is a declining function of radius. This of course assumes that the surface density profiles do not evolve in time. One source of clumps in the tail is hence expected to be stripped ISM gas, if the ram-pressure exceeds the criterion in Eq.~\eqref{GunnGott}. The functional forms of our stellar and gas surface density ($\Sigma_{\star\text{,\,disc}}$ and $\Sigma_\text{gas,\,disc}$, respectively) can be found in \citet{2023arXiv231105679S}. There, we even find that the inner parts of the disc at radii smaller than the scale radius are stripped during the central passage in the cluster.

\subsection{Galaxy formation physics}\label{Methods:simulations}

We use the Auriga model of galaxy formation \citep{2017MNRAS.467..179G} with the two main differences that we do not use active galactic nucleus feedback, and we use a fixed wind velocity, which is only realistic for a MW mass galaxy. We use the ISM model from \citet{2003MNRAS.339..289S}, where gas cells with a density higher than the number density threshold for star formation, $n_\text{sf}=0.157$ cm$^{-3}$, are multi-phase with a contribution from a cold and a hot phase. Stellar population particles are probabilistically spawned from the star-forming cells, and stellar winds are released according to stellar evolution models (see \citealt{2013MNRAS.436.3031V} for details).

\subsection{Post-processing: ionisation modelling}

We use the CLOUDY \citep{1998PASP..110..761F,2017RMxAA..53..385F} tables presented in \citet{2018MNRAS.475.1160H} to calculate the \iona{H}{i} content of our gas cells. We take into account a uniform UV background (from \citealt{2009ApJ...703.1416F}), radiative cooling, self-shielding, and collisional ionisation.

As mentioned in Section~\ref{Methods:simulations}, our ISM model treats star-forming gas cells as having a subgrid contribution from cold and hot phases. In our post-processing analysis we set the temperature of star-forming gas cells to $10^3$ K. Our results are not sensitive to this value, since our main purpose is to ensure that the star-forming gas is included in the ISM according to the criterion in Sect.~\ref{Sec.:GasReservoirs1}. An identical approach is often used in the literature, for example, in \citet{2023MNRAS.518.5754R}.

\subsection{Our definitions of ISM, CGM, and ICM}\label{Sec.:GasReservoirs1}

We now turn to the task of selecting gas residing in the ISM, CGM, and ICM. We base our definition on a combination of the thermal properties and the spatial location of a gas cell. We define the centre of the galaxy, ($x_\text{galaxy},y_\text{galaxy},z_\text{galaxy}$), to be the location of the most bound particle or cell in the simulation. As a measure of the radial extent of the gas of the \iona{H}{i} disc ($R_\text{HI}$), we determine the distance from the galaxy centre to the upstream location where the disc drops below a characteristic HI column density of $10^{20.5}$ cm$^{-2}$ (column density maps and profiles in Appendix~\ref{AppendixRHI} motivate this choice). At times earlier than 2.60 Gyr, HI clouds are abundant in the CGM upstream of the galaxy. In order not to bias our disc estimates, we model the average column density profile with a parabola and perform a fit to determine $R_\text{HI}$. At later times, the HI clouds in the CGM upstream of the galaxy centre are stripped and the HI column density is monotonically increasing towards the galaxy centre. Thus, we can simply identify the extend of the HI disc by determining the point furthest upstream with a column density above $10^{20.5}$ cm$^{-2}$. We further demonstrate and validate our methods for determining $R_\text{HI}$ in Appendix~\ref{AppendixRHI}.

To characterise gas in different phases, we use the following definition of the ISM, CGM, and ICM gas reservoirs:\\
\noindent{}{\bf ISM}: Gas with $T\leq 10^{4.5}$ K and no more than $1.1 R_\text{HI}$ downstream from the galaxy centre ($y\leq y_\text{galaxy}+1.1 R_\text{HI}$). We multiply $R_\text{HI}$ with a factor of 1.1 to bias our selection towards being inclusive when defining gas in the ISM. The spatial requirement ensures that the ISM belongs to the disc and not to clumps in the jellyfish tail far downstream from the galaxy.\\
\noindent{}{\bf ICM}: Gas with $\log( T/\text{K}) \geq 0.2 \log[ \rho /(\text{M}_\odot~ \text{kpc}^{-3})] + 5.8$. This density dependent temperature cut is successful in isolating the thermal properties of our simulation's injection region (the ICM) from the CGM.\\
\noindent{}{\bf CGM}: Everything else.\\
In Appendix~\ref{appendix1} we motivate these selections by analysing the density-temperature diagram.

During infall, the stripping radius, which we identify with the smallest radius where the inequality in Eq.~\eqref{GunnGott} is valid, is smaller in comparison to $1.1 R_\text{HI}$. This is consistent with the jellyfish galaxy's ISM experiencing ram-pressure stripping at these times. The stripping radius shrinks from 5.9 kpc (as the galaxy enters the cluster and passes through $R_{200}$) to 1.0 kpc (which is reached at the centre of the cluster). In the same time interval, $1.1 R_\text{HI}$ shrinks from 33.0 kpc to 8.0 kpc.

\subsection{Post-processing: friends-of-friends clump finding}

We use a friends-of-friends (FoF) algorithm to find groups of gas cells with an \iona{H}{i} number density larger than $0.01$ cm$^{-3}$ that are at a distance larger than $1.1 R_\text{HI}$ downstream from the galaxy centre. The density threshold of $0.01$ cm$^{-3}$ corresponds to an FoF linking length of 0.62 kpc (calculated as described in section~3.4 of \citealt{2023arXiv231105679S}), and we require a minimum of 10 cells in an FoF group. We use a serial python code as described further in \citet{2023arXiv231105679S}. We refer to the identified `FoF groups' as either `tail clumps' or simply `clumps' throughout this paper.

\begin{figure}
\centering
\includegraphics[width=0.95\linewidth]{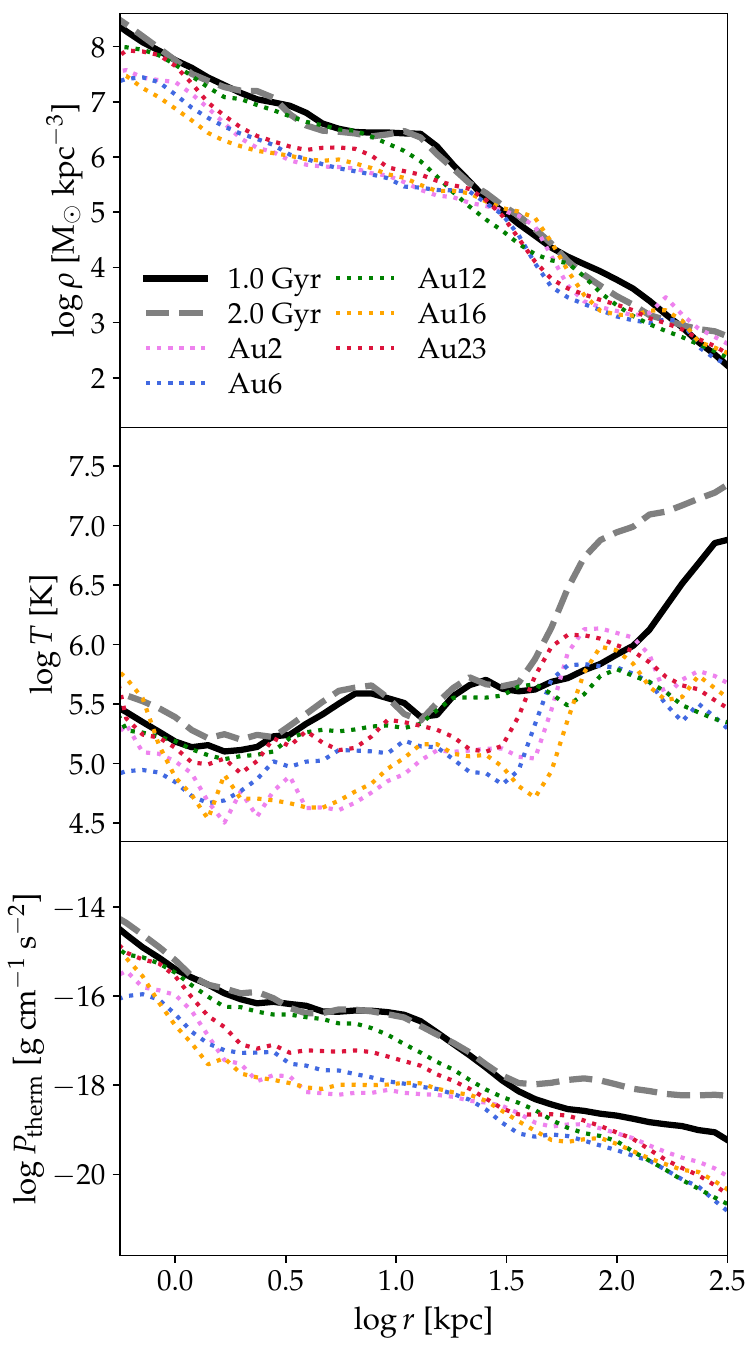}
\caption{Density, temperature, and thermal pressure as a function of radius for our simulated jellyfish galaxy at 1.0 and 2.0 Gyr. We compare to the redshift $z=0$ profiles of five different simulations from the Auriga suite. Note that the increase of temperature and thermal pressure at the largest radii delineates the transition from the CGM to the hot ICM wind. Hence, there is a good agreement between our simulation and Auriga, implying that our adopted CGM is consistent with cosmological simulations.}
\label{Analyse17A_PlotDensityProfiles_denstiy_T_plot}
\end{figure}

\subsection{Tracing of gas}

To track the Lagrangian path of gas mass we use Monte Carlo tracer particles \citep{2013MNRAS.435.1426G}. At $t=0$ we create five tracers within each gas cell. During the simulation tracer particles are transferred between gas cells, stellar population particles, and wind particles, such that an ensemble of tracers enables us to follow Lagrangian gas flows in a statistical fashion. The use of tracer particles (instead of directly following the paths of gas cells) is correct and preferred, since there is mass transfer between different gas cells.

First, we focus our tracer analysis at a time of 2.50~Gyr. We use our FoF algorithm to identify \iona{H}{i} clumps in the tail, downstream from the jellyfish galaxy. For each tail clump, we select all the tracer particles and track their density, temperature and spatial coordinates back in time. Based on their properties at $t=1$ Gyr we determine their origin to be the ISM, CGM or ICM. This time is chosen, because it corresponds to the time the `burn-in phase' has finished and the galaxy is in a quasi-equilibrium. We use the notation $\text{CGM}\to\text{clumps}$ to refer to gas tracers, which started out in the CGM at $t=1$ Gyr and reside in tail clumps at $t=2.50$~Gyr. We use a similar notation for tracers starting in the ISM and ending up in tail clumps, see Table~\ref{Table:gastracers}.

For the ICM we introduce the tag `$\text{ICM}\to\text{clumps}$', for all tracers, which at $t=1$~Gyr had a density and temperature obeying our ICM-criterion or at a time of $1.0\leq t/\text{Gyr}\leq 2.50$ resided in the injection region of the simulation windtunnel.

For the tracers characterised as `$\text{CGM}\to\text{clumps}$', we define a subset of tracers, which have had an in-between phase residing in the disc's ISM: this is refereed to as $\text{CGM}\to \text{ISM}\to \text{clumps}$. Similarly, we introduce the notation $\text{ICM}\to \text{ISM}\to \text{clumps}$ for ICM tracers that have been accreted onto the ISM before ending up in the galaxy tail.

If we let $f$ denote the fractional contribution to the \iona{H}{i} mass in a clump, we can calculate the fraction of gas, which has resided in the ISM at some point prior to a time of $2.50$~Gyr as
\begin{align}
f(\text{through ISM}) =& f(\text{ISM}\to\text{clumps})\notag \\&+f(\text{CGM}\to \text{ISM}\to\text{clumps})\notag\\&+f(\text{ICM}\to \text{ISM}\to\text{clumps}).\label{eq:fthroughISM}
\end{align}
This quantity will play a key role in our analysis of gas flows. In this subsection, we have explained how our tracer analysis was used to characterise gas at a snapshot of 2.50~Gyr. We repeat this process at times of 2.75 and 3.25 Gyr, and present our results in the next section.

\begin{figure*}
\centering
\includegraphics[width=\linewidth]{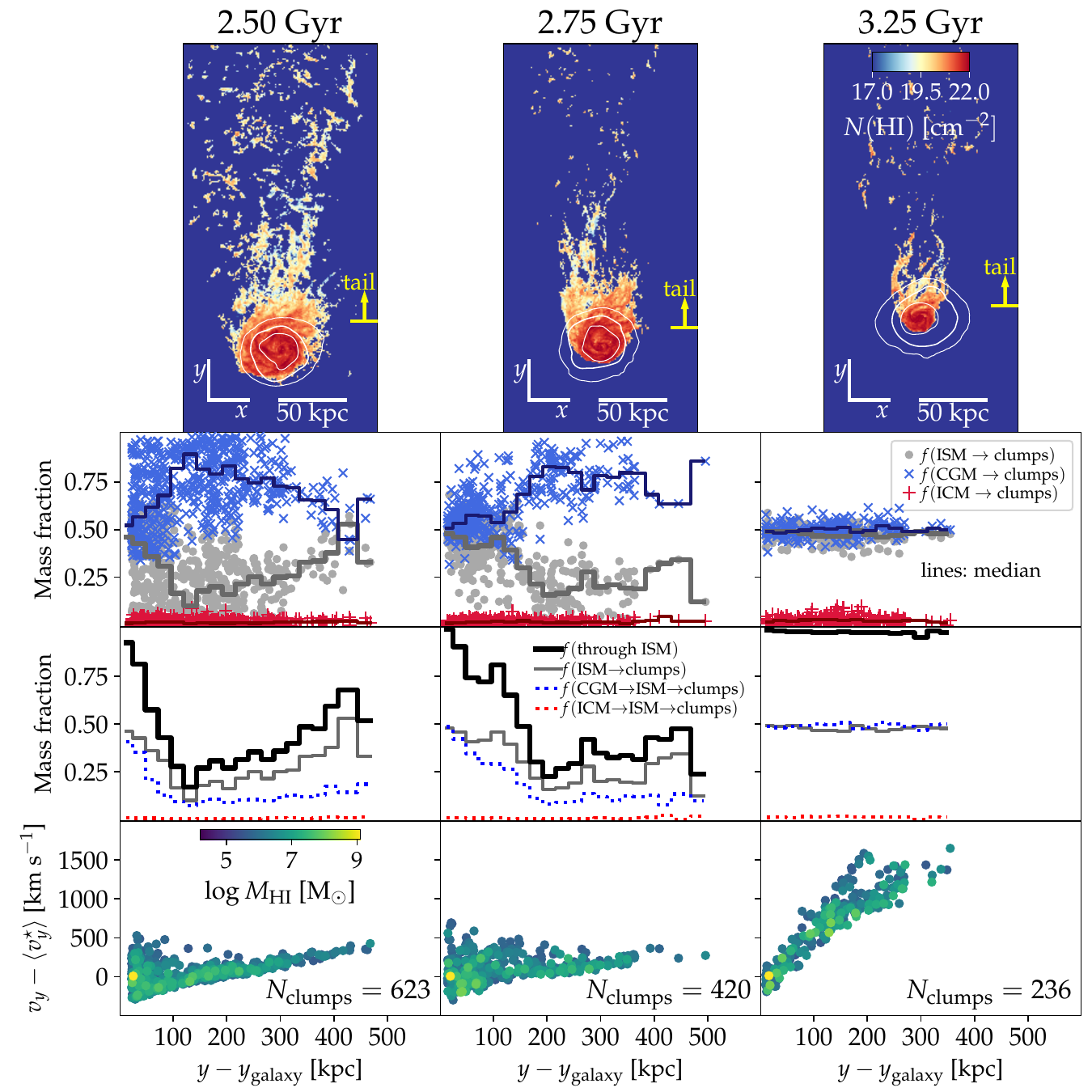}
\caption{\iona{H}{i} column density (upper panels), mass fraction of tail clumps initially residing in the ISM, CGM or ICM (second row), mass fraction of different paths through the ISM from Table~\ref{Table:gastracers} (third row), as well as the mass-weighed $\varv_y$-velocity of each tail clump (lower panels) for three times (2.50, 2.75, and 3.25 Gyr so that the central passage takes place in between the second and third analysed time). The majority of mass in the tail clumps initially resided in the CGM for these three snapshots. Well before the central passage (at 2.50 Gyr) the dominant mass fraction of the tail's \iona{H}{i} clumps are accreted directly from the CGM. After the central passage (at 3.25 Gyr), about half of the gas in the tail initially resided in the CGM, but almost exclusively has been accreted onto the ISM before it got stripped and finally assembled in the jellyfish tail. The other half of the gas originates in the initial ISM that was present already at $t=1$ Gyr.}
\label{Analyse12_FoF_Shattering_withtracers_letterversion1Gyr_testsim2HR_RotMatrix30deg_095_TracerAnalysis_2}
\end{figure*}

\section{Results}\label{Sec:Results}

\subsection{Validating the thermal profiles with state-of-the-art cosmological simulations}

Before analysing the role of CGM gas in the build up of the jellyfish tail, we will validate the initial CGM of the jellyfish galaxy. To do this we analyse our simulation after the burn-in phase at 1.0 Gyr and at 2.0 Gyr, when the galaxy enters $R_{200}$ of the cluster. In this validation we compare our CGM profiles to five galaxies from the cosmological Auriga simulations \citep{2017MNRAS.467..179G} of Milky Way mass galaxies at a redshift of $z=0$.

In Fig.~\ref{Analyse17A_PlotDensityProfiles_denstiy_T_plot}, we plot the thermal profiles calculated in spherical shells equally spaced in logarithmic radius ($\log r$).  For $r\gtrsim 100$ kpc, our galaxy is significantly hotter than the Auriga simulations. The reason is the hot ICM wind. Except for this physically desired effect, we find quantitatively good agreement between our simulation and Auriga implying that we have a CGM consistent with state-of-the-art cosmological simulations of MW-mass galaxies. Our initial conditions are hence ideal for studying how ram-pressure stripping affects the CGM.

The galaxy experiences an ICM density of $10^{2.7}$~M$_\odot$~kpc$^{-3}$ as it enters $R_{200}$, and a peak of $10^{4.7}$~M$_\odot$~kpc$^{-3}$ at the time of the nearest passage of the cluster centre (see Fig.~\ref{Fig10_PlotClusterWindForPaper2}). The CGM on the other hand (see radii ranging from $\sim10^{1.5}$~kpc to $10^2$~kpc in Fig.~\ref{Analyse17A_PlotDensityProfiles_denstiy_T_plot}) spans densities ranging from $10^4$ to $10^5$~M$_\odot$~kpc$^{-3}$. Around the time of the infall, the CGM is hence in the cloud-crushing regime ($\rho_\text{CGM}>\rho_\text{ICM}$) as identified by \citet{2024arXiv240402035G}, and closer to the nearest passage there would co-exist zones in the cloud-crushing- and bubble-regimes (the latter regime is defined as $\rho_\text{CGM}<\rho_\text{ICM}$) -- which is a realistic scenario as discussed in section~5.2.1 of \citet{2024arXiv240402035G}.

\begin{figure}
\centering
\includegraphics[width=\linewidth]{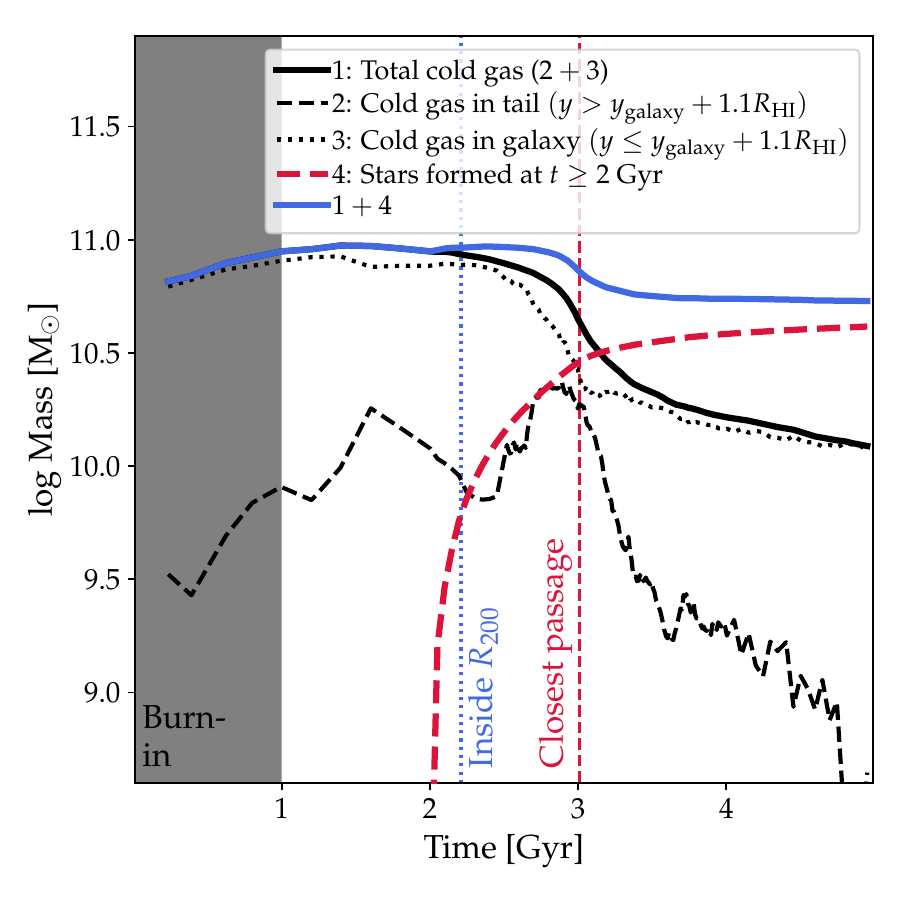}
\caption{Time evolution of the total cold gas mass (defined as $T\leq 10^{4.5}$ K) in the simulation (1), the cold gas in the tail (2), cold gas associated with the galactic disc (3), and the total mass stars formed since $t=2$ Gyr (4). As expected from idealised simulations of cloud-wind interactions in the regime of large clouds, the cold gas mass of the tail increases after entering the cluster until a time close to the closest central passage. Continued star formation and enhanced ram-pressure stripping of the ISM during the central passage causes a drop of the cold gas in the disc and tail.}
\label{tmpAnalyse8_MassEvolution_March11ChristophCommentEvolution.pdf}
\end{figure}

\subsection{The origin of gas in the tail}\label{subsec:gasintails}

We now turn to the key question of this paper and analyse the tail of the jellyfish galaxy. We present the analysis for three different times: 2.50, 2.75, and 3.25 Gyr. The first two times sample the infall of the galaxy onto the galaxy cluster before the central passage, and the latter time is after the central passage.

In Fig.~\ref{Analyse12_FoF_Shattering_withtracers_letterversion1Gyr_testsim2HR_RotMatrix30deg_095_TracerAnalysis_2} we plot the \iona{H}{i} column density (upper row) to present a visualisation of the cold gas in the tail in the wake of the galaxy, which points in the direction of the ICM velocity (and which is also found in the cosmological simulations of \citealt{2019MNRAS.483.1042Y}). The white contours indicate the surface density of stars. The yellow arrow marks the $y$-value, which divides the tail from the galactic disc ($y_\text{galaxy}+1.1R_\text{HI}$). For the tail clumps identified at each snapshot, we show the mass fraction of the origin of the cold gas (either $\text{ISM}\to\text{clumps}$, $\text{CGM}\to\text{clumps}$, or $\text{ICM}\to\text{clumps}$) in the second row. The mass fraction of the gas that at some point has been a part of the cold ISM according to Eq.~\eqref{eq:fthroughISM} is shown in the third row, together with the sub-classes outlined in Table~\ref{Table:gastracers}. The mass-weighted $y$-velocity for each clump relative to the stars in the galaxy, that is $\varv_y -\langle \varv_y^\star \rangle $, is shown in the fourth row. Note that the velocity of the ICM wind relative to the galaxy ($\varv_\text{wind}-\langle \varv^\star_y\rangle$) is 2211, 2591 and 2771 km s$^{-1}$ for a time of 2.50, 2.75, and 3.25 Gyr, respectively. Hence, the tail clumps are not sufficiently accelerated to be co-moving with the wind.

We see a remarkable \iona{H}{i} tail present at all three times with a clumpy structure characteristic of shattering as we concluded in \citet{2023arXiv231105679S}. The downstream clumps are dominated by gas, which initially resided in the CGM; especially at $t=2.50$ and $2.75$ Gyr. This has two reasons. First, a fraction of the CGM gas is directly pushed by ICM ram pressure into the tail without ever entering the disc's ISM. This can, for example, be seen $\gtrsim 200$ kpc downstream at 2.75 Gyr, where $f(\text{CGM}\to \text{clumps})$ is significantly larger than $f($through ISM$)$. Secondly, the gas in our simulated galaxy (and galaxies in general, see \citealt{2019MNRAS.483.4040S}) is continuously being transferred between the CGM and ISM. This highlights the result, that it is necessary to include a realistic CGM in idealised jellyfish galaxy simulations to reliably study the structure of the tail.

We now further assess the fraction of gas, $f$(through ISM), which at some point during the simulation has been accreted to the disc's ISM before entering the shattered tail. There is a transition occurring, when comparing the times before (2.50 and 2.75 Gyr) and after (3.25 Gyr) the central passage. At the latter time, the clear majority of gas in the shattered tail has resided in the ISM at some point during the simulation while at the same time, about 50 per cent of this gas originated in the CGM (i.e. the channel $\text{CGM}\to \text{ISM}\to \text{clumps}$ slightly dominates). At the two earlier times, a lower fraction of gas (in comparison to 3.25 Gyr) resided in the ISM before entering the tail. Thus, during the galaxy's infall there is a larger fraction of gas, which has never been part of the ISM. This gas has been mixed with stripped cold ISM to an intermediate temperature that is thermally unstable, so that it lost its thermal energy via radiative cooling before it enters the cold phase in the tail and shatters into smaller cloudlets.

The majority of the cold gas in the tail originates in the CGM, which is less dense and hence much easier to accelerate by the wind in comparison to the more dense ISM. Hence, this explains the enormous extend of some jellyfish tails. Mixing in hot ICM and cooling it to $10^4$ K further adds momentum to the tails (in the galaxy rest frame), but this phase is small by mass in comparison to the other two phases.

This transition is also visible in the \iona{H}{i} projections. At 3.25 Gyr the clumps in the tail are elongated, which is a consequence of these clumps having been directly stripped from the ISM, in accordance with the classic picture of ram-pressure stripping \citep{1972ApJ...176....1G}. At earlier times, however, the clumps are less elongated and `more shattered' (following the nomenclature of \citealt{2018MNRAS.473.5407M}), so here thermal instability and radiative cooling of CGM gas, which has never resided in the ISM, is more important.

The finding that the `through-ISM-channel' is completely dominating after the time of the nearest cluster passage is of course a result of the ram-pressure stripping history. To some extend, this is a coincidence of the choice of our adopted galaxy orbit that implies this transition to occur at around the closest passage. 

The monotonically increasing trend in $\varv_y$ versus $y-y_\text{galaxy}$ shows that clumps are accelerated as they move downstream. Such a trend is expected during the infall in the cluster. We see that the tail gas moves faster downstream after the central passage in comparison to before. So after the passage we do not only see a transition in the origin of the gas, but also in the tail velocity.

The above results are insensitive to the choice of FoF parameters. We varied the minimum number of cells that constitute a clump between 5, 10 (our standard choice), and 20. We also ran the analysis with a 10 times high FoF density threshold in comparison to our standard value ($0.01$ cm$^{-2}$). These variations of course changed the total number of FoF groups and their masses, but they did not change the overall findings and scalings of the curves in Fig.~\ref{Analyse12_FoF_Shattering_withtracers_letterversion1Gyr_testsim2HR_RotMatrix30deg_095_TracerAnalysis_2}.

Using our intermediate resolution simulations from \citet{2023arXiv231105679S}, we have also checked how our Fig.~\ref{Analyse12_FoF_Shattering_withtracers_letterversion1Gyr_testsim2HR_RotMatrix30deg_095_TracerAnalysis_2} would change if the angle between the ICM wind and the galaxy's rotation axis would change from 60$^\circ$ (as in the simulation presented in this paper) to 0, 30, and 90$^\circ$. Our most important results, which is that $f($through ISM$)$ increases for the tail later in the simulation, is independent of this angle. That being said, the experienced ram-pressure is of course inclination-dependent, as we also discuss in \citet{2023arXiv231105679S}.

\begin{figure*}
\centering
\includegraphics[width=\linewidth]{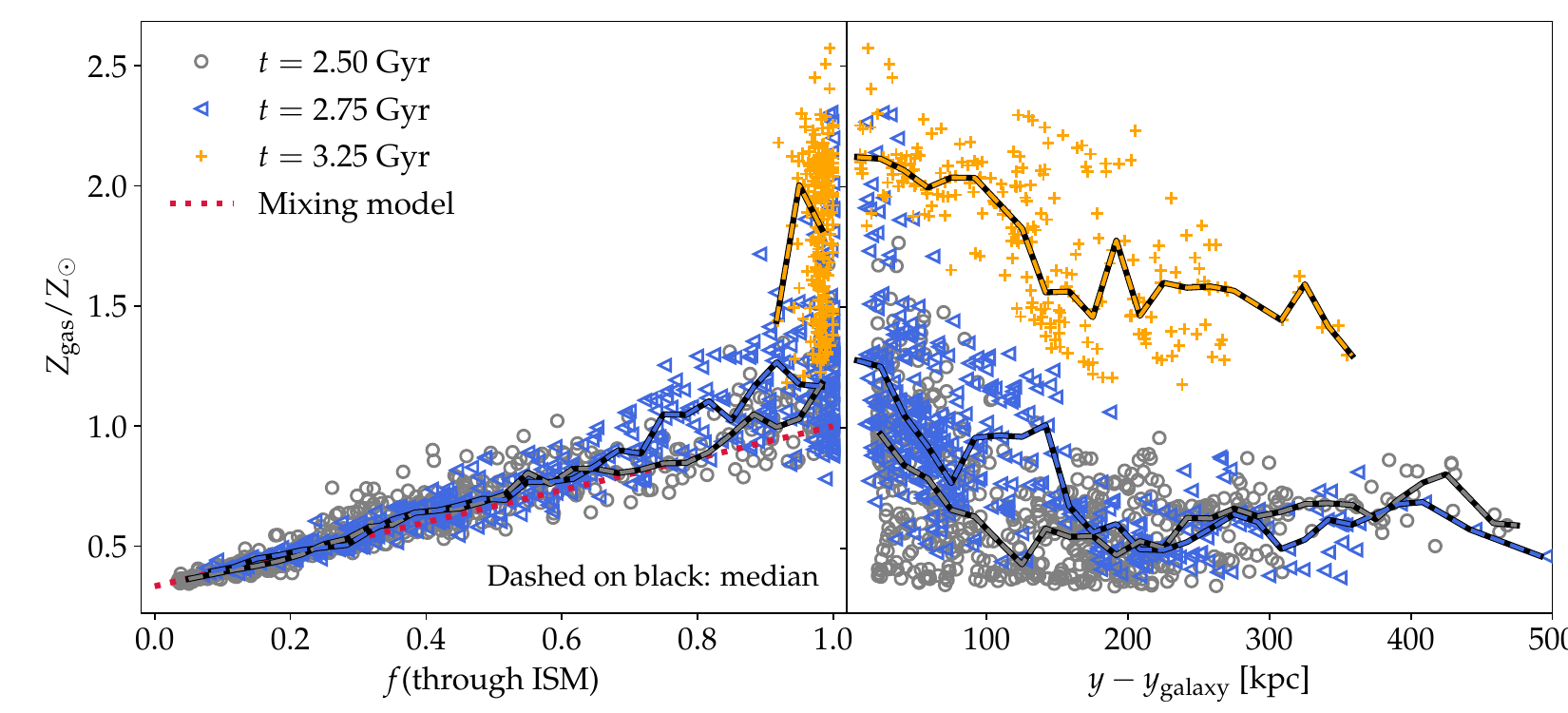}
\caption{Gas metallicity of the clumps in the tail, which are shown as different symbols denoting the different analysis times. Left panel: We show the gas metallicity as a function of the mass fraction of gas, which at some point during the simulation has resided in the jellyfish galaxy's ISM. During infall (2.50 and 2.75 Gyr), there is a correlation between these two quantities. This indicates that the metallicity of the jellyfish tail is influenced by mixing of high-metallicity ISM gas with low-metallicity gas from CGM and ICM. We quantify this with a `mixing model', $Z/\text{Z}_\odot=\frac{2}{3}f\text{(through ISM)}+\frac{1}{3}$, which assumes that the ISM metallicity does not evolve. After the closest passage in the cluster (3.25 Gyr), the clear majority of gas in tail clumps has resided in the ISM during the simulation, and we see a higher metallicity than predicted by the non-evolving mixing model. Right panel: We plot the metallicity as a function of distance downstream from the galaxy, and we find a declining trend after the central passage. Before the central passage we also see a decreasing trend in the immediate tail within $\lesssim 80$ kpc of the galaxy. This is a signature that mixing is more important further downstream in comparison to close vicinity of the galaxy. }
\label{Metallicity_Analyse12_FoF_Shattering_withtracers_testsim2_RotMatrix30deg}
\end{figure*}

\subsection{The baryonic mass budget}

In Fig.~\ref{tmpAnalyse8_MassEvolution_March11ChristophCommentEvolution.pdf} we assess the build-up of the tail from a mass budget perspective. We focus on the cold gas mass with a temperature below $10^{4.5}$~K. We plot the total cold gas mass in the simulation (1), the cold gas mass in the tail (2), the cold gas mass associated with the galaxy (3), and the stars formed after $t=2$~Gyr (4), which is when the galaxy started to experience an appreciable ram pressure. The mass of cold gas in the galactic disc significantly declines from the entrance of $R_{200}$ until the central passage in the cluster. This is an effect of ram-pressure stripping, as well as stars forming during the infall (also shown in the figure). Conversely, the mass of cold gas in the tail grows by a factor of three from the infall to the central passage. This is expected from idealised simulations of cloud-wind interactions in the `growth' regime of large clouds \citep[e.g.][]{2017MNRAS.470..114A, 2018MNRAS.480L.111G, 2020MNRAS.499.4261S, 2020MNRAS.492.1841L}. Continued star formation and enhanced ram-pressure stripping of the ISM during and after the central passage, furthermore, causes the amount of cold gas to decline in the disc and tail. Note, that the formation of stars at late times is consistent with the finding from \citet{2023MNRAS.524.3502R} that the jellyfish galaxies continue forming stars until they have lost approximately 98 per cent of their gas. Overall, the evolution of the gas shows the evolution from a gas-rich spiral galaxy to a less star-forming, gas poor galaxy during the first infall onto the cluster. This is also consistent with \citet{2023MNRAS.524.3502R} who find that ram-pressure stripping is capable of stripping jellyfish galaxies from $z=0.5$ to $z=0$.

\subsection{The metallicity in the gaseous tail}

The physics of jellyfish tails can, for example, be constrained by radio observations \citep{2021NatAs...5..159M,2021A&A...650A.111R,2022ApJ...924...64I}, but a promising observable is also the gas metallicity of the clumps. \citet{2021ApJ...922L...6F} has shown that the gas metallicity in the tail is influenced by the mixing of gas from the ICM wind and the galaxy. Here we further study this connection.

We remind the reader that we initialise gas in the ISM with solar metallicity, $Z_\text{ISM}=\text{Z}_\odot$, and gas in the CGM and ICM is initialised with a three times lower metallicity, $Z_\text{CGM,ICM}=\text{Z}_\odot/3$. In Fig.~\ref{Metallicity_Analyse12_FoF_Shattering_withtracers_testsim2_RotMatrix30deg} (left panel), we show the gas metallicity of tail clumps as a function of $f$(through ISM). During the infall in the cluster (at 2.50 and 2.75 Gyr) there is a strong correlation, which is expected because an over-abundance of metals is initialised and produced in the galactic ISM. After infall (at 3.25 Gyr), the majority of gas in the tail clumps has resided in the ISM at some point during the simulation. In the meanwhile, the ISM has been significantly enriched during the simulation time so that the tail is not only characterised by a larger metallicity in comparison to earlier times but has also super-solar metallicities.

We adopt a simple model for the jellyfish tail metallicity:
\begin{align}
Z_\text{tail}=(Z_\text{ISM}-Z_\text{CGM,ICM})f\text{(through ISM)}+Z_\text{CGM,ICM}.
\end{align}
Specified to our simulations, this model assumes that all gas, which ever was in the galactic ISM has solar metallicity and everything else has $1/3$ solar, so that it assumes that the ISM metallicity is not evolving. This toy model has the implicit assumption that the inflow of low-metallicity gas into the ISM and the outflow rate of enriched gas are balanced by metals produced by star formation. We refer to this as a `mixing model'.

This model correctly predicts an increasing metallicity as a function of $f\text{(through ISM)}$ (observed at times 2.50 and 2.75 Gyr) and delineates the lower envelope of the relation. This is expected because the model does not account for additional enrichment due to ongoing star formation and stellar mass return and indicates that mixing of ISM and CGM gas can explain the metallicity of our tail clumps. At a time of 3.25 Gyr, the ISM has been significantly enriched, so here the model no longer gives an accurate description.

In the right panel of Fig.~\ref{Metallicity_Analyse12_FoF_Shattering_withtracers_testsim2_RotMatrix30deg}, we show the gas metallicity as function of the downstream distance from the galaxy. At all times, we see a declining metallicity as a function of distance; at $t=2.50$ and $2.75$ Gyr the metallicity is higher in the immediate tail at $y-y_\text{galaxy}\lesssim 80$ kpc in comparison to further downstream, and at $t=3.25$ Gyr we see a more prevalent monotonically declining trend reaching further downstream from the galaxy. We thus overall, confirm the declining metallicity in the tail as seen in \citet{2021ApJ...922L...6F}, and also in the recent modelling of \citealt{2024arXiv240402035G} (see their sect. 4.1.3).

\section{Discussion on ISM and CGM stripping}\label{Sec:Discussion}

The CGM gets completely stripped in our simulation. This can be seen by comparing our Fig.~\ref{MovieHR2024Appendix_010__12Apr_Movie} and Fig.~\ref{MovieHR2024Appendix_045__12Apr_Movie}, which show the HI column density at a time of
2 Gyr and 2.75 Gyr, respectively. At the latter time, there is a complete lack of $>10^{19}$ cm$^{-2}$ gas upstream from the disc. On a similar note, a short stripping time-scale of CGM gas was found in satellites orbiting a MW-like halo in \citet{2024arXiv240400129Z}, who found a CGM stripping time-scale of a few hundred Myr. We note that we have a peak ram pressure of approximately $3\times 10^{-10}$ g cm$^{-1}$ s$^{-2}$, which is three orders of magnitudes larger than \citet{2024arXiv240400129Z} ($3.94\times 10^{-13}$ g cm$^{-1}$ s$^{-2}$), but they also simulate a less-massive system, which explains the similarly short stripping time-scales. \citet{2024arXiv240400129Z} also found a large fraction of the ISM to survive the ram-pressure stripping, similar to us (see fig. 2 of \citealt{2023arXiv231105679S}).

\citet{2024MNRAS.527..265R} did a similar analysis of ISM stripping time-scales in simulations of satellites orbiting a MW-like galaxy. They found ISM stripping time-scales of 0.25 to 3-4 Gyr, depending on the satellite's mass. The ISM of their most-massive satellite was continuously fed by gas being accreted from the host MW-mass halo's reservoir. Although we do not see gas being accreted from the ICM to the ISM as a dominant channel, we find the similar result that ram pressure stripping in our simulations is not capable of fully removing the ISM of the jellyfish galaxy.

In this paper, we have not attempted to resolve the star-forming clouds or the ISM, and instead we have imposed the star formation law from \citet{2003MNRAS.339..289S}. The ISM-associated observables are therefore not necessarily correctly predicted. We therefore leave it to future work, to carry out simulations with a resolved ISM model and to predict the above-mentioned observables. Expanding our simulation setup to a more structured ISM modelling for a jellyfish galaxy orbiting a galaxy cluster will likely reveal the impact of the existence of these two tail formation mechanisms, and it may even lead to an evolutionary sequence for jellyfish galaxies and their tails. A natural follow-up question is also how the signatures of the two mechanisms are imprinted into the common jellyfish observables, such as maps of UV intensity, emission lines, X-ray, synchrotron/radio and molecular gas.

Another advantage of using a structured ISM model is associated with the mixing layer, which has a temperature in-between the stripped cool gas and the hot gas \citep{2021ApJ...911...68T,2024MNRAS.527..265R}. This cooling may induce formation of condensed clouds \citep{2018MNRAS.480L.111G,2023arXiv230703228A}. For this to happen, it is, of course, necessary to fully capture the density and temperature gradients of the gas in the mixing layer surrounding the cold gas. A potential caveat here, comes from the treatment of the ISM in our physics model. When gas is denser than the star formation threshold, it is described by an effective equation of state and stellar winds are probabilistically ejected from the gas cells. The stellar wind particles are decoupled from the hydrodynamical interactions until they are outside the star-forming region. Because of this behaviour, we do not expect the detailed structure of the star-forming gas or the layers surrounding them to be fully realistic in our current simulations. Rerunning with a structured ISM model that models galactic winds emerging from a combination of supernova, cosmic ray and radiative energy injection would improve the modelling of the mixing layers and the ISM structure \citep{2024arXiv240513121T}.

\section{Conclusion}\label{Sec:Conclusion}

This paper has revealed two distinct formation mechanisms of the tails of jellyfish galaxies:

\begin{enumerate}
\item When ram-pressure is not sufficiently strong to strip the majority of the ISM, tails are formed by CGM gas being transferred into the tail where they mix with the stripped ISM to intermediate temperatures. Those are subject to thermal instability so that the ensuing fast radiative cooling causes the formation of multi-phase tails.
\item When the ram pressure from the ICM is strong enough, the ISM is directly stripped and moved into the wake of the galaxy, where a dense tail is formed. This mechanisms follows the idea of the analytical model of \citet{1972ApJ...176....1G}.
\end{enumerate}
These two formation mechanisms are clearly identified in our numerical simulation, but it is still an open question whether and how these can be identified in observations.

A very recent publication by \citet{2024arXiv240402035G} also identifies the importance of including the CGM, when studying the stripping of jellyfish galaxies and the build-up of their tails. In comparison to us, they do a comprehensive study of stripping regimes in multiple simulations, where they identify the `bubble-mode' and `cloud-crushing-regime' for CGM stripping. The strategy of our paper is to include a time-varying wind in a single simulation, and also include radiative cooling processes, star-formation, and chemical enrichment, and this enables us to study, for example, the metallicity evolution and shattering of tail clumps. We find it encouraging that both studies identifies the CGM to play a key role in forming jellyfish tails.

\begin{acknowledgements}
We acknowledge support by the European Research Council under ERC-AdG grant PICOGAL-101019746.
\end{acknowledgements}

%
\bibliographystyle{aa} 
\bibliography{ref2}

%

\begin{appendix}

\section{Defining the ISM, CGM and ICM gas reservoirs}\label{appendix1}

\begin{figure}[h]
\centering
\includegraphics[width=\linewidth]{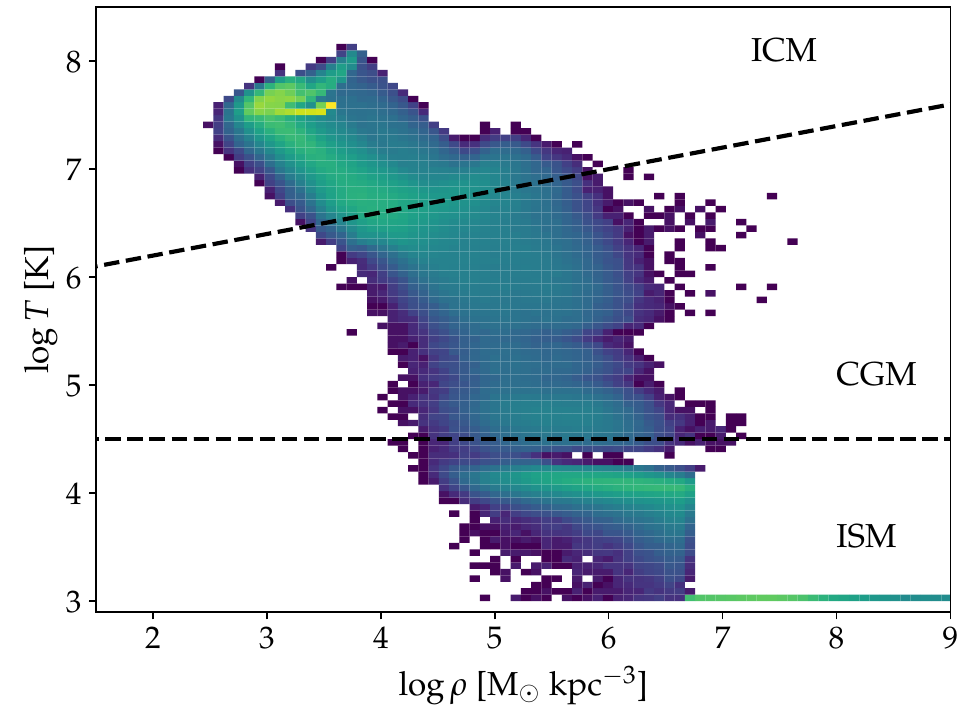}
\caption{Tracer classification based on the $(\rho,T)$-plane. The background histogram shows the gas cells at a time of $2.50$ Gyr, and the dashed lines divide our definitions of ISM, CGM, and ICM. In Sect.~\ref{Sec.:GasReservoirs1} we show quantitative definitions of these reservoirs.}
\label{RhoT_Identification_Analyse12_FoF_Shattering_withtracers_testsim2_RotMatrix30deg}
\end{figure}

We now describe our choices of the ISM, CGM, and ICM definition from Sect.~\ref{Sec.:GasReservoirs1} in detail. With these definitions the $(\rho,T)$-plane is used to divide up the phases, according to the dashed lines in Fig.~\ref{RhoT_Identification_Analyse12_FoF_Shattering_withtracers_testsim2_RotMatrix30deg}. We have adopted an ISM definition aimed at including cold gas in the disc. We have adopted a definition of ICM gas selecting all gas in the wind injection region during the simulation. Finally the CGM is defined to be gas which is neither ISM or ICM -- this requirement effectively selects halo gas colder or comparable to the virial temperature. Our definitions are explicitly stated in Sect.~\ref{Sec.:GasReservoirs1}.
\FloatBarrier
\FloatBarrier

\section{Determination of the HI disc radius}\label{AppendixRHI}

We now introduce and define the HI disc radius ($R_\text{HI}$), which we use in Section~\ref{sec:methods} to define the tail of the jellyfish galaxy. Conceptually, we would like to determine it as the radius, where the HI column density of the disc reaches the value typically associated with a galaxy's ISM ($\approx 10^{20.5}$--$10^{22}$~cm$^{-2}$, see table~5 and fig.~8 of \citealt{2017MNRAS.469.2959K}).

\begin{figure*}[h]
\centering
\begin{minipage}{0.45\textwidth}
\centering
\includegraphics[width=\linewidth]{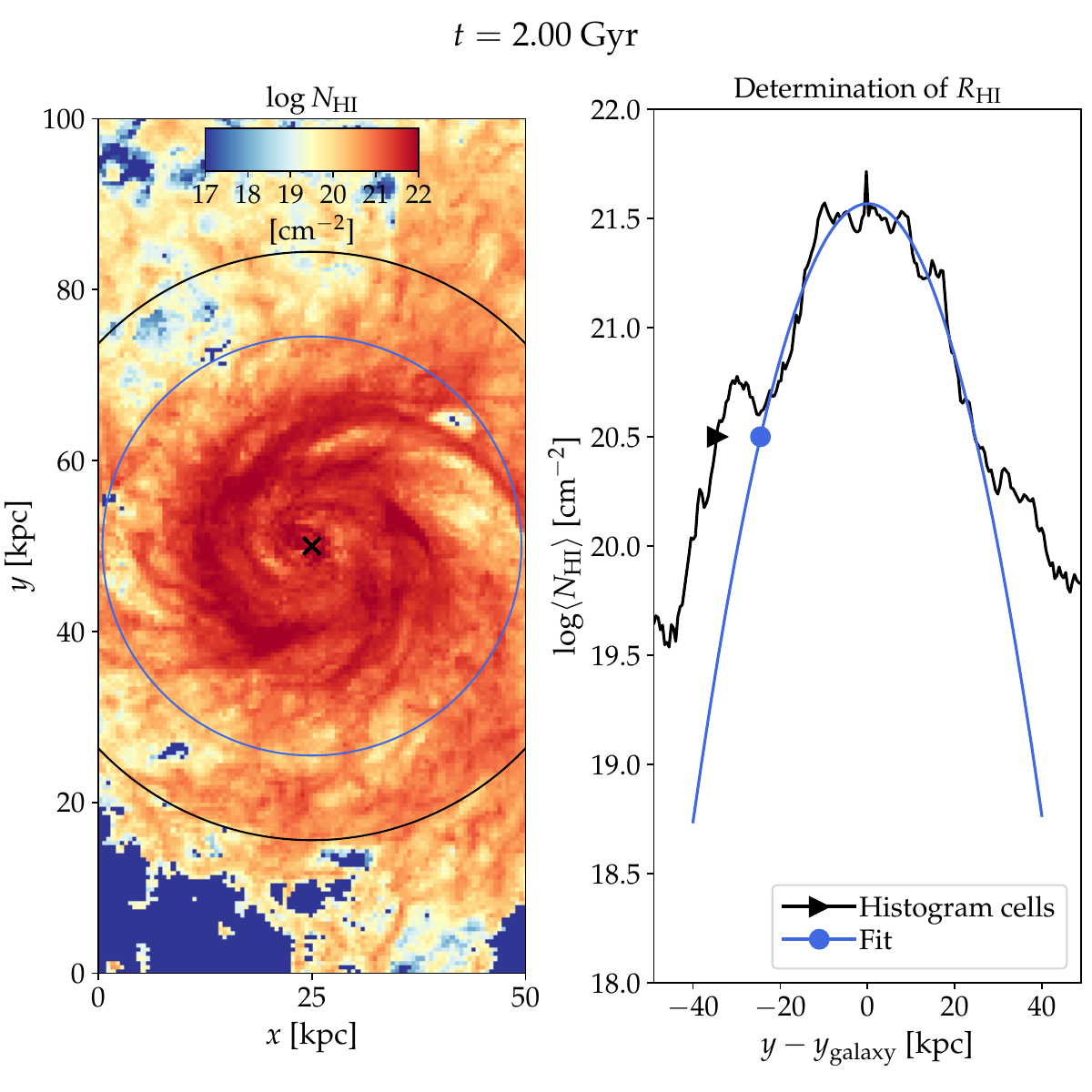}
\caption{HI projected column density of the jellyfish galaxy (left panel), where the cross marks the galaxy centre, and the blue triangle and black circle mark $R_\text{HI}$ determined with a fit and by identifying the furthest upstream location with $\langle N_{\rm HI}\rangle\geq 10^{20.5}$ cm$^{-2}$, respectively. In the right panel, we plot $\langle N_{\rm HI}\rangle $ (see text for details) and mark $R_\text{HI}$ determined with the different methods. At this early time of $t=2.0$ Gyr, there are CGM clumps upstream from the galaxy, so we use a fit to determine the $R_\text{HI}$ for the disc.}
\label{MovieHR2024Appendix_010__12Apr_Movie}
\end{minipage}
\hspace{0.02\textwidth}
\begin{minipage}{0.45\textwidth}
\centering
\includegraphics[width=\linewidth]{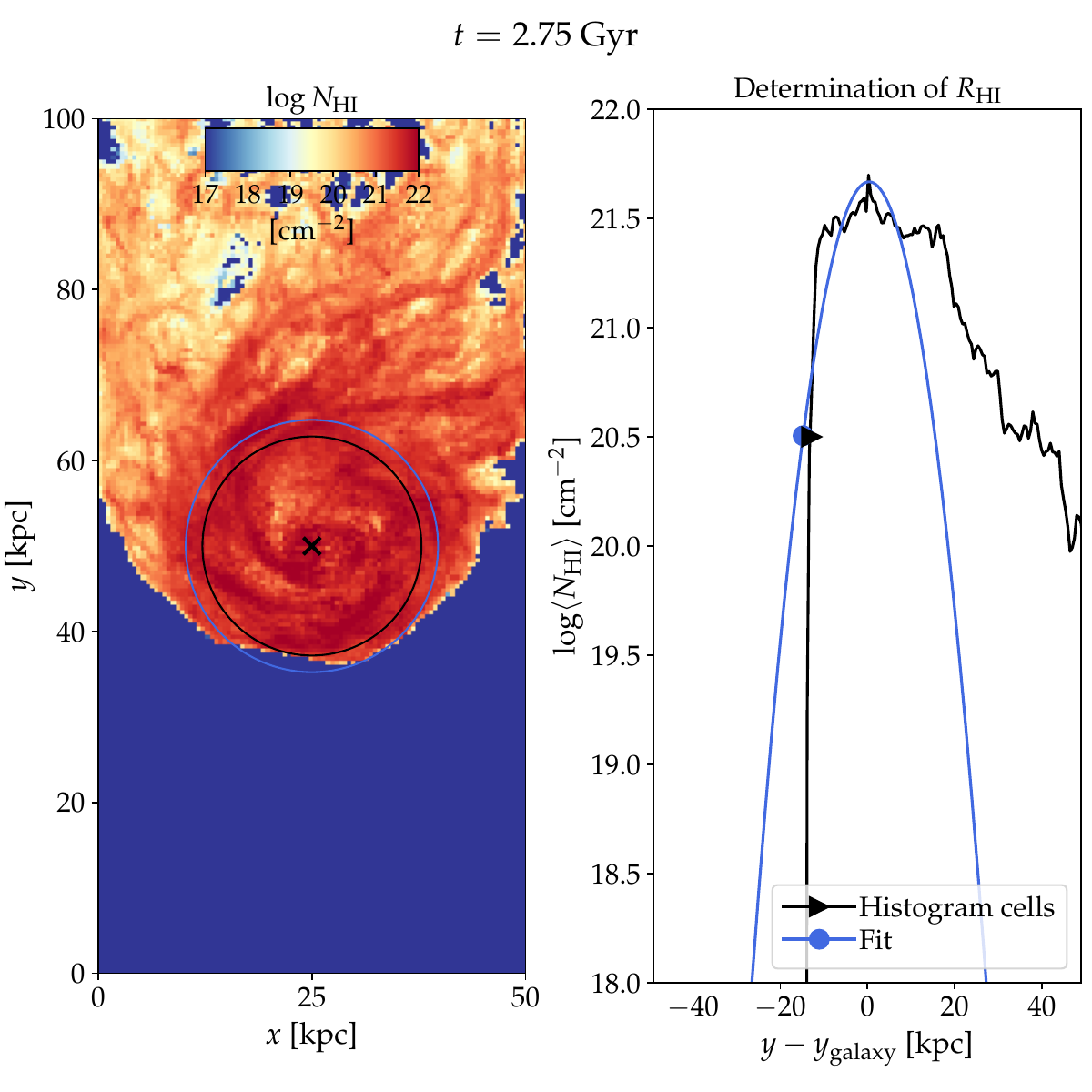}
\caption{Same as Fig.~\ref{MovieHR2024Appendix_010__12Apr_Movie}, but for a time of $t=2.75$ Gyr, where we instead of fitting the averaged column density with a parabola simply determine the furthest upstream location with $\langle N_{\rm HI}\rangle\geq 10^{20.5}$~cm$^{-2}$ (see black triangle). This approach is sufficient and preferred at times, where HI clumps in the CGM upstream the galaxy are absent.}
\label{MovieHR2024Appendix_045__12Apr_Movie}
\vspace{1.75em}
\end{minipage}
\end{figure*}

We determine the projected HI column density in the $(x,y)$-plane using a $100\times 200$ histogram grid distributed at $x-x_\text{galaxy}=\pm 25$ kpc and $y-y_\text{galaxy}=\pm 50$ kpc. For each of these grid cells, we calculate $N_\text{HI}\equiv \frac{1}{150 \text{ kpc}}\int_{z-z_\text{galaxy}=-75 \text{ kpc}}^{75 \text{ kpc}} n_\text{HI} \, {\rm d}z$ numerically. The resulting projection plots at a time of 2.0 and 2.75 Gyr are shown in Fig.~\ref{MovieHR2024Appendix_010__12Apr_Movie} and \ref{MovieHR2024Appendix_045__12Apr_Movie}, respectively (left panels). Next we determine the characteristic HI value as a function of $y$, by averaging over $x$: $\langle N_\text{HI}\rangle \equiv \frac{1}{50 \text{ kpc}}\int_{x-x_\text{galaxy}=-25 \text{ kpc}}^{25 \text{ kpc}} N_\text{HI} \, {\rm d}x $. The right panels of Fig.~\ref{MovieHR2024Appendix_010__12Apr_Movie} and \ref{MovieHR2024Appendix_045__12Apr_Movie} show this averaged profile.

By comparing the HI projection at 2.0 and 2.75 Gyr, we see that the former snapshot has a richer and more inhomogeneous gaseous structure upstream from the galaxy. This is a contribution from the ISM and CGM, which has not yet been stripped at 2.0 Gyr. To derive an extent of the HI disc, we here fit a second order polynomial of the form, $\log \langle N_\text{HI} \rangle = A (y-y_\text{galaxy})^2 + B (y-y_\text{galaxy}) + C$. This fit is shown in the right panel of the figures. At 2.0 Gyr we use this fit to determine the $y$-value, where the disc HI column density reaches a value of $10^{20.5}$ cm$^{-2}$.

\begin{figure}
\centering
\includegraphics[width=\linewidth]{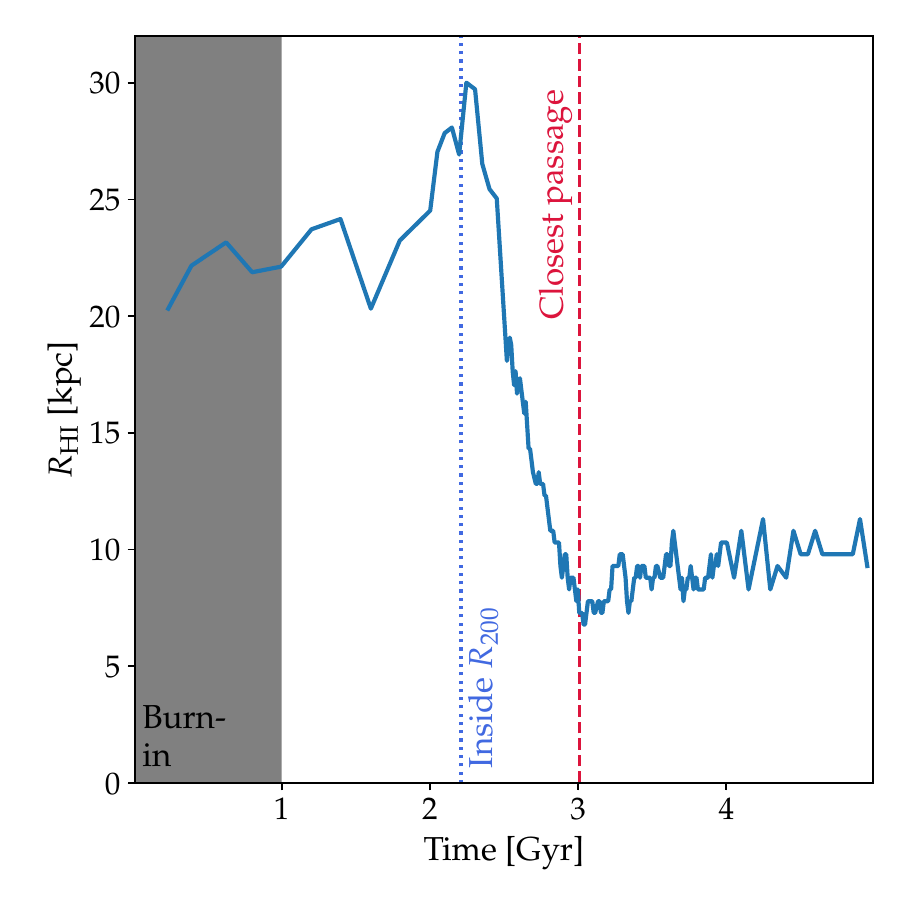}
\caption{Evolution of the HI radius, which has been determined with a fit for $t\leq 2.60$ Gyr, and a simple detection of the furthest upstream location with $\langle N_{\rm HI}\rangle\geq 10^{20.5}$~cm$^{-2}$ at later times.}
\label{Analyse999_RHIEvolution}
\end{figure}

At later times, for example, at 2.75 Gyr, the CGM upstream from the disc has been stripped. Instead of performing a fit, we simply take the first $y$-value with $\langle N_\text{HI}\rangle \geq 10^{20.5}$ cm$^{-2}$. At $t\leq 2.60$ Gyr we perform fits as described above, and at later times, we identify the lowest $y$-value with $\langle N_\text{HI}\rangle \geq 10^{20.5}$ cm$^{-2}$. We visually inspect all snapshots of the simulations to check that this gives a reasonable behaviour of the HI disc radius. In the left panel of Fig.~\ref{MovieHR2024Appendix_010__12Apr_Movie} and \ref{MovieHR2024Appendix_045__12Apr_Movie} we plot the fitted- and value-identified radius as a blue and black circle, respectively. The time evolution of our identified $R_\text{HI}$ is shown in Fig.~\ref{Analyse999_RHIEvolution}. We see that this quantity increases from the start of the simulation until the galaxy enters $R_{200}$ in the galaxy cluster. This is because of gas accreted to the ISM from the CGM. As the ram-pressure increases the HI radius declines until the nearest passage. At later times, the HI radius slowly increases in our simulation. The resulting time-depending HI radius enables us to classify, whether gas downstream from the galaxy centre belongs to the tail or the disc ISM (Sect.~\ref{Sec.:GasReservoirs1}).

\end{appendix}

\end{document}